\documentclass[conference]{IEEEtran}

\IEEEoverridecommandlockouts
\usepackage{cite}
\usepackage{amsmath,amssymb,amsfonts}
\usepackage{algorithmic}
\usepackage{algorithm}
\usepackage{graphicx}
\usepackage{textcomp}
\usepackage[dvipsnames]{xcolor}
\usepackage{comment}  
\usepackage{url}

\usepackage{tikz}
\usetikzlibrary{calc, arrows, shapes, chains}
\usetikzlibrary{decorations.pathreplacing}
\usetikzlibrary{positioning, fit}
\usetikzlibrary{automata}

\definecolor{ballblue}{rgb}{0.13, 0.55, 0.95}
\tikzstyle{polygon}=[very thick, draw=ballblue]
\tikzstyle{point}=[ultra thick, draw=gray, cross out, inner sep=0pt,minimum width=4pt,minimum height=4pt,]
\tikzstyle{line}=[red]
\tikzset{MyArrow/.style={single arrow, draw=black, minimum width=10mm, minimum height=18mm, inner sep=2mm, single arrow head extend=1mm}}

\def\BibTeX{{\rm B\kern-.05em{\sc i\kern-.025em b}\kern-.08em
    T\kern-.1667em\lower.7ex\hbox{E}\kern-.125emX}}
\begin{document}

\title{A Case for Network-wide Orchestration of Host-based Intrusion Detection and Response}

\author{\IEEEauthorblockN{1\textsuperscript{st} Mark Timmons}
\IEEEauthorblockA{\textit{Department of Computer Science} \\
\textit{Naval Postgraduate School}\\
Monterey, CA \\
mark.timmons@nps.edu}
\and
\IEEEauthorblockN{2\textsuperscript{nd} Daniel Lukaszewski}
\IEEEauthorblockA{\textit{Department of Computer Science} \\
\textit{Naval Postgraduate School}\\
Monterey, CA \\
dflukasz@nps.edu}
\and
\IEEEauthorblockN{3\textsuperscript{rd} Geoffrey Xie}
\IEEEauthorblockA{\textit{Department of Computer Science} \\
\textit{Naval Postgraduate School}\\
Monterey, CA \\
xie@nps.edu}
}
\IEEEspecialpapernotice{This work has been submitted to the IEEE for possible publication. Copyright may be transferred without notice, after which this version may no longer be accessible.}
\maketitle

\begin{abstract}
Recent cyber incidents and the push for zero trust security underscore the necessity of monitoring host-level events. However, current host-level intrusion detection systems (IDS) lack the ability to correlate alerts and coordinate a network-wide response in real time. 

Motivated by advances in system-level extensions free of rebooting and network-wide orchestration of host actions, we propose using a central IDS orchestrator to remotely program the logic of each host IDS and collect the alerts generated in real time. In this paper, we make arguments for such a system concept and provide a high level design of the main system components. Furthermore, we have developed a system prototype and evaluated it using two experimental scenarios rooted from real-world attacks. The evaluation results show that the host-based IDS orchestration system is able to defend against the attacks effectively. \\

\end{abstract}


\section{Introduction}

Host-based intrusion detection and response not only complements but also offers a distinct advantage over traditional network-based defenses: 
Unlike perimeter-focused security mechanisms, a host-based solution operates closer to the point of attack, enabling real-time analysis of application behavior and early mitigation of attacks~\cite{han20,ribeiro20}. One such mitigation may include early detection of a malicious URL~\cite{anderson16} from packets traversing a host's protocol stack, and then intervention to thwart the connection at the host.  Additionally, with the adoption of zero trust access control~\cite{nist_zerotrust}, network operation increasingly relies on host-based IDSs.  

Meanwhile, there are solutions such as Linux kernel modules and eBPF~\cite{ebpf-book} to extend kernel level security functionality without having to reboot the host. Furthermore, for hosts operating as virtual machines, virtual machine introspection (VMI)~\cite{garfinkel2003,simplifyingVMI} can be used to create an independent entity for monitoring and even modifying host behavior at the granularity of system calls. Leveraging this host level programmability, a recent effort~\cite{lukaszewski2023AgileNetOps} developed a network-wide orchestration system, called Layer 4.5, to remotely program and coordinate actions at multiple hosts to enhance network programmability. 

Following a similar line of thinking, we hypothesize that network-wide orchestration of host-based IDS actions is not only feasible, but also offers new functionality that is not possible with today's IDS deployment. Suppose a central controller, which we will refer to as \emph{an IDS orchestrator}, is able to remotely program the logic of each host-based IDS and collect the alerts generated in real time. The orchestrator will then be able to start a multi-round, multi-fidelity \emph{interrogation of host behavior}. An interrogation that is triggered by an alert on one application process or one traffic flow can lead to the deployment of additional logic from the orchestrator to look more closely at some system properties or perform the required mitigation. Additionally, the orchestrator should be able to correlate alerts over time and across each host IDS when assessing a new threat, which can help reduce false positives and increase the effectiveness of threat response.

 As a motivating scenario, consider a known attack where criminal and advanced persistent threat actors use the BlackEnergy malware toolkit~\cite{MITRE_S0089} to create botnets for denial-of-service attacks and hijack benign host applications as camouflage to establish connections with malicious command-and-control servers. An IDS orchestrator performing host behavior interrogation upon suspicious application behaviors can  detect these malicious processes with high confidence and then enact an automatic response.

This paper takes a first step towards understanding the power and limitation of network-wide orchestration of host IDS actions. Our contributions are three-fold as follows: 
\begin{enumerate}
\setlength\itemsep{0.55em}
    \item We have formulated a system concept for the orchestration, including a high level design of the main system components. 
    
    \item We have developed a system prototype based on the open-source Layer 4.5 codebase~\cite{github}. 
    
    \item We have evaluated the system prototype using two experimental scenarios motivated by real-world attacks. The results show that the prototyped orchestration was able to defend against the attacks effectively. 
\end{enumerate}

The rest of the paper is organized as follows.  In Section II, we present the details of our system concept and prototype implementation. Sections III and IV describe our experimental design and results, respectively.  We discuss potential system extensions and related work in the next two sections (V and VI). Finally, we conclude the paper with Section VII. 

\nocite{han20,ribeiro20}

\section{Orchestration of Host-based IDS}

In this section, we present a system for orchestrating IDS actions across edge devices that include hosts, servers and middleboxes. For simplicity, except in some situations where we need to distinguish the device type, we'll refer to them generally as hosts in the rest of the paper.

\subsection{System Concept}\label{concept}

Central to our system is an IDS orchestrator that requires a secure communications channel to all hosts that have a corresponding IDS agent installed.
Furthermore, we require that the orchestrator has the capacity to remotely update the functionality of a host IDS agent without rebooting the host.

Our system aims to provide the following two new capabilities by leveraging the network-wide view and direct control of the IDS orchestrator. We believe that these capabilities are not possible today, with host actions alone.
 
\begin{itemize}
  \item {\bf Multi-fidelity flow interrogation}. An interrogation that is triggered by one flow can expand to additional flows and hosts. The same can be said about the types of conditions to examine. Being able to dynamically create and install new IDS functionalities on select hosts, the orchestrator is well-equipped to carry out such iterative interrogations in a timely manner.

  \item {\bf Network-wide event correlation}. The correlation can be both temporal (i.e., over time) and spatial (i.e., across hosts). Being able to calibrate observations from many vantage points and over long durations, the orchestrator is well-equipped to differentiate truly malicious activities from those benign one-off deviations in traffic behavior.  
    
\end{itemize}

\begin{figure} [t]
    \centering
    \includegraphics[width=0.8\linewidth, trim=0 2 2 2, clip]{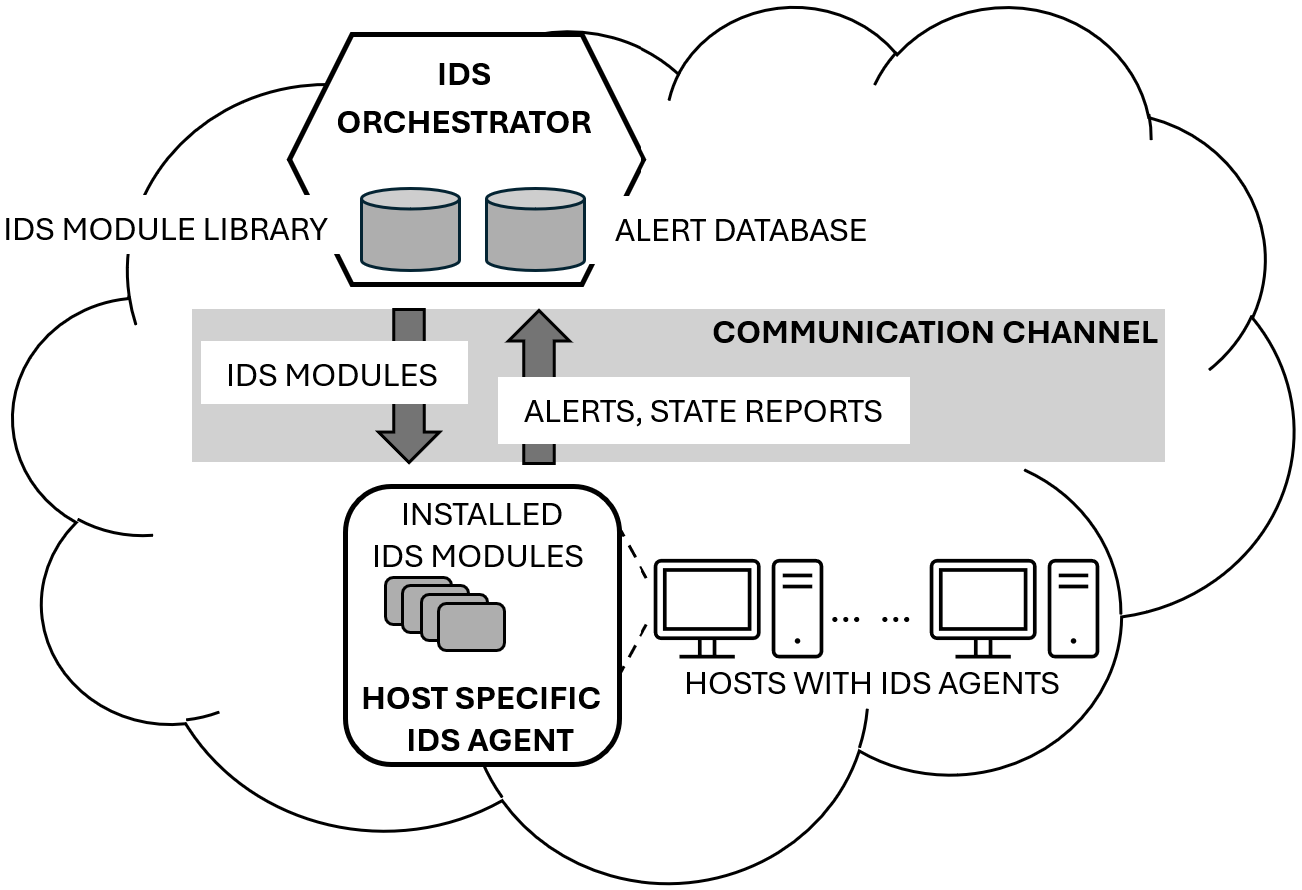}
    \caption{Illustration of main design components of proposed system. 
    }
    \label{fig:Network_Orch}
\end{figure}

Next, we provide more design details about the three main components of the proposed orchestration system, as illustrated in Figure~\ref{fig:Network_Orch}.

\subsubsection{IDS Orchestrator}
The IDS orchestrator serves as a central decision maker within the network. Upon receiving alerts of adverse conditions from one or more host IDS agents, the orchestrator determines an appropriate response and, if necessary, sends additional modules to all host IDS agents that need to participate in the response. The response can be further host interrogations with the newly deployed modules, or throttling of specific network traffic to stop further spread of malicious activities. 

\subsubsection{Host IDS Agent} 
A host IDS agent monitors the behavior of a host for malicious activity.  It consists of a collection of software modules, called IDS modules, that are remotely installed by the orchestrator. The modules gather information from the application processes and each layer of the protocol stack to enable a granular approach to intrusion detection, which minimizes false positive rates while providing key data for IDS orchestrator event correlation. 
Some modules will be able to perform certain actions autonomously based on the detected threat, sharing such actions with the orchestrator afterwards. 

\subsubsection{IDS Module Deployment}
IDS orchestrator response is based on the contextual information provided by the host agent alerts.  
One such response involves deploying additional IDS modules, which can be accomplished through a combination of pre-built or Just-in-Time (JIT) dynamically generated modules.

Pre-built modules perform binary decision making to detect the presence of an alerting condition.  These types of modules act based on known signatures with the purpose of alerting the orchestrator upon detection of a possible attack and taking standard response actions.  This alert triggers IDS orchestrator deployment of a ``next-tier'' module providing additional host scrutiny.
As flow interrogation proceeds, the need for more dynamic modules will occur, which we term as JIT modules.  The key difference between module types, is that JIT modules require up to date host IDS agent information prior to deployment.  For instance, after a pre-built module detects a known attack signature, the orchestrator begins collecting data via host interrogation.  The host information gathered from the increased scrutiny is then incorporated into applicable host response modules for construction and deployment.

A multi-tiered defense in depth approach to intrusion detection and response would utilize both types of modules to provide a diverse strategy to defeat known and novel attack vectors \cite{defense_in_depth}.
To achieve this defensive posture, the IDS orchestrator needs automated processes to quickly parse through the host-generated log data and build the next iteration of response modules.  Incorporating deep reinforcement learning techniques—proven effective in complex cybersecurity tasks~\cite{nguyen21}—offers significant advantages for the orchestrator.

\subsection{Prototype Implementation}
There are multiple methods available today to conduct security monitoring on the host, such as Virtual Machine Inspection (VMI)~\cite{simplifyingVMI}, eBPF~\cite{ebpf-book}, and the Layer 4.5 framework~\cite{lukaszewski2023AgileNetOps}.  The VMI approach allows for interrogating the virtual host processes from the physical host.  This method should be more resistant to root kit attacks, but requires all hosts to run as VMs.  The Cilium project~\cite{cilium} uses the eBPF method to improve the observability, security, and networking of container clusters.  This method would allow demonstrating some orchestration functionality, but would require all the host functions to be containerized. Of note, the available code bases for both the VMI and eBPF projects are lacking a network-wide orchestrator component. 

In comparison, the Layer 4.5 framework supports deployment of general purpose kernel modules via a network orchestrator~\cite{lukaszewski2023AgileNetOps} and its code base is open source~\cite{github}.  Additionally, the framework allows for the dynamic replacement of kernel modules on existing kernel flows without interruption.  For this reason, we chose to leverage the Layer 4.5 framework as a starting point for our prototype.  In the following, we first describe the modifications made to the Layer 4.5 system and then identify the system limitations because of this choice.

\subsubsection{Orchestrator Modification}
\begin{figure}[htb]
    \centering
        \resizebox{50mm}{!}{
         \includegraphics[width=0.95\linewidth]{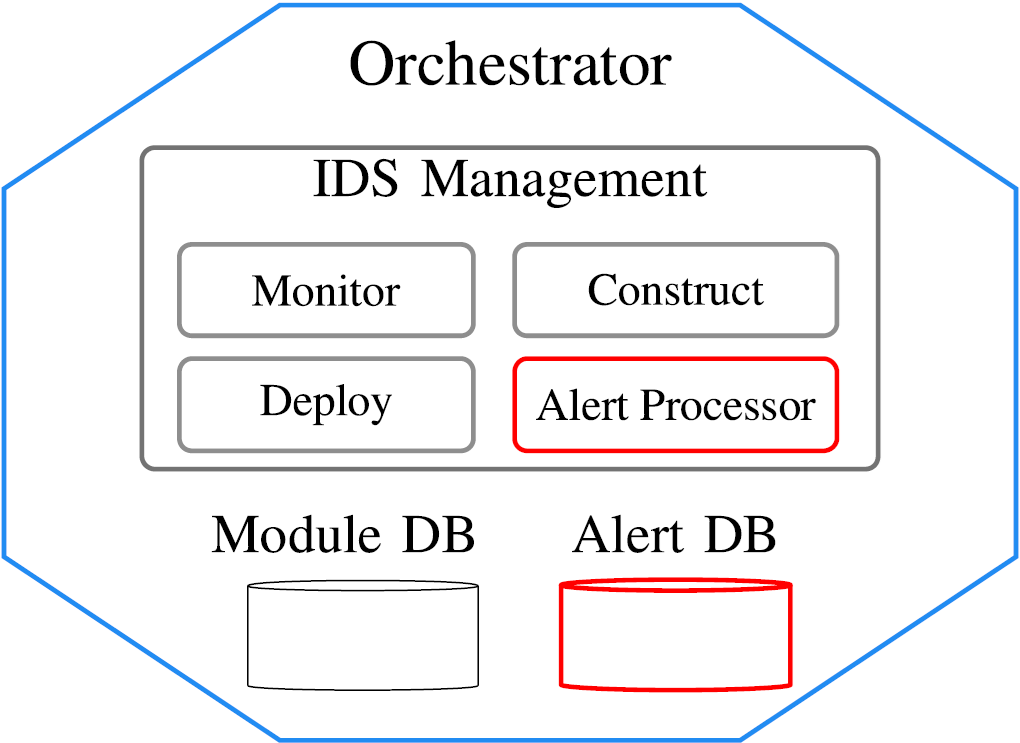}
    }

    \caption[Layer 4.5 Orchestrator.]{Adapted from \cite{lukaszewski2023AgileNetOps}. Modified Layer 4.5 orchestrator consists of two new components: Alert Processor method and Alert Database (in red).}
    \label{fig:nco}
\end{figure}

The Layer 4.5 framework~\cite{lukaszewski2023AgileNetOps} includes a network orchestrator that establishes a control channel with Layer 4.5 compatible network devices to monitor and deploy customization modules.   We adapt the orchestrator to function as an IDS orchestrator in Figure \ref{fig:nco}. The IDS orchestrator leverages the existing monitor, construct, and deploy functions, but we now include an ``Alert Processor" method for handling IDS module alerts and the ability to store the provided alert conditions within an alert database.




\subsubsection{Alert Processing}
Algorithm \ref{alg:alert} provides the core logic added to the IDS orchestrator to handle host IDS agent alerts. The IDS orchestrator begins by monitoring all host connections, ready to process any new alerts.  When a new alert is received, the orchestrator first stores the alert condition in the alert database (3).  After the alert information is stored, the orchestrator determines if the alert condition triggers the required deployment of a new IDS module (4).  If a new module is required, the orchestrator first determines which hosts within the network will receive the new IDS module, regardless of the current alert status on that host (5).  Once the list of hosts has been determined, the orchestrator determines if the new module was pre-built or if JIT construction is required (6).  
As previously discussed, JIT construction is necessary if the new IDS module requires contextual alert information to properly build the module.
If JIT construction is required, then the orchestrator gathers alert information from the alert database (7) and then uses that information to build (8) the module for each applicable host.  Finally, the orchestrator deploys (9) the new IDS module throughout the network. 

\begin{algorithm}
\caption{Orchestrator: Alert Processor}\label{alg:alert}
\begin{algorithmic}[1]
\STATE Monitor for alert $A$ from IDS module $M$
\STATE Upon receiving $A$:
\STATE \hspace{3mm} Record $A$ in alert database;
\STATE \hspace{3mm} \textbf{if} $A$ triggers next-tier IDS module $M_2$ \textbf{then}
\STATE \hspace{8mm} Determine hosts (set $H$) requiring new agent $M_2$;
\STATE \hspace{8mm} \textbf{if} JIT construction is required \textbf{then}
\STATE \hspace{12mm} Collect data for $M_2$ construction;
\STATE \hspace{12mm} Build $M_2$ for each $h \in H$;
\STATE \hspace{8mm} Deploy $M_2$ to each $h \in H$;
\end{algorithmic}
\end{algorithm}

\subsubsection{Prototype Limitations}\label{limitations}
By leveraging capabilities from the Layer 4.5 framework~\cite{lukaszewski2023AgileNetOps}, our current prototype has some limitations on what system concepts can be quickly accomplished. Overcoming these limitations is outside the scope of this work.  

The first limitation is that Layer 4.5 requires an established TCP or UDP socket for the modules to perform intrusion detection because it only supports inspecting data in these socket buffers. Therefore, it does not monitor the entire network stack and any process that utilizes other means for attack, such as RAW sockets, will not be detected by our IDS modules. 

Second, to minimize changes to Layer 4.5, we utilized the existing monitoring, construction, and deployment methods in Figure \ref{fig:nco}. This decision results in sub-optimal orchestrator-host response times.  To reduce the time between alert generation and orchestration notification without inducing significant control channel overhead, we set the monitor interval to 5 seconds instead of the standard 30 seconds.
\section{Design of Experiments}
    In this section, we begin by describing known attack vectors identified in the MITRE ATT\&CK framework, which drive the design and objectives of our experiments.  Next we provide a description of the testbed, the tests performed, and the criteria to determine experiment success or failure.    All the open sourced code to reproduce the experiments within this paper is available on GitHub~\cite{github} (included prior to publication).

\subsection{Attack Vectors and Objectives}\label{attacks}

The first attack technique targets Network Denial of Service, specifically a Domain Name Service (DNS) flood attack~\cite{MITRE_T1498}. A DNS flood attack from a host or multiple hosts will attempt to overwhelm the target DNS server with a large number of queries. Mitigation for the attack includes intercepting the malicious traffic upstream of the affected service, or blocking the attacking source IP address and protocol locally. 

The second attack technique targets application-layer protocols, enabling attackers to introduce tools or establish command-and-control communications with an external server~\cite{MITRE_T1105}. 
A common response is to use intrusion prevention systems with appropriate signatures, along with protocol-based filtering techniques, to detect and block adversary traffic~\cite{MITRE_T1071}.  When data flows are encrypted using TLS or SSL, detecting malicious socket data in clear text may not be possible. However, analyzing contextual information within network flows has proven effective in identifying malicious traffic and mitigating the impact of encryption~\cite{anderson16}.

Using these two attack scenarios, we aim to fulfill three main objectives.  First, the initial IDS module must be able to detect the attack and successfully alert the IDS orchestrator.  Second, the network-wide orchestration must automatically process and respond by deploying a new IDS module. Last, the new module must be successfully installed on the host IDS agent to stop/mitigate the attack.

\subsection{Testbed Design}
\begin{figure}[htb]
    \centering
    \includegraphics[trim={1 5 1 5}, clip, width=0.9\linewidth]{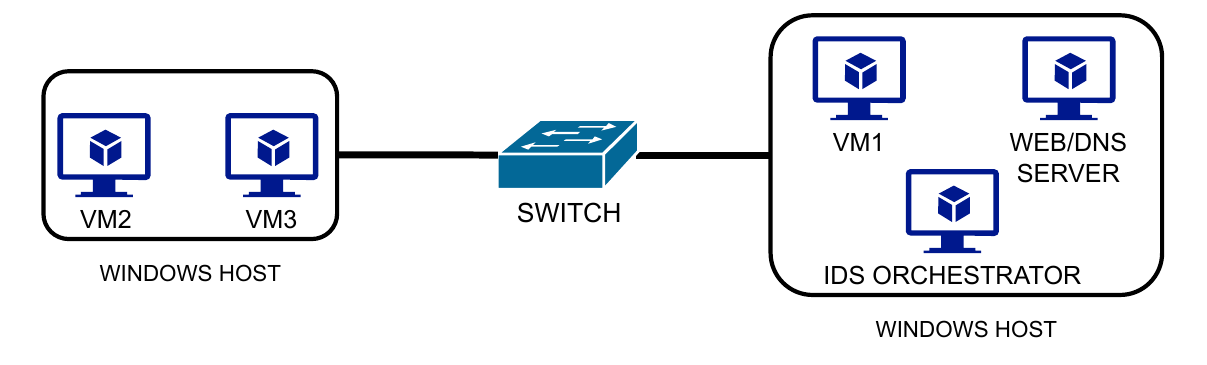}
    \caption{Experimental testbed consisting of two physical hosts running multiple virtual machines.}
    \label{fig: Network Topology}
\end{figure}

The network architecture used for experimentation, shown in Figure~\ref{fig: Network Topology}, creates a realistic networking environment that demonstrates connectivity between multiple endpoint devices. 
This architecture mimics a small enterprise network, showcasing the ability to detect and mitigate intrusions both from a network-wide perspective and at the host level.
We chose a relatively small testbed topology to focus on a functional evaluation of the orchestrator's ability to respond to intrusions. As previous work~\cite{lukaszewski2023AgileNetOps} has shown, we anticipate that a typical two-level hierarchy of access and backbone components would result in a moderate increase in orchestrator response time.   

The testbed includes two Windows 11 laptops, each running Ubuntu 20.04 Layer 4.5 enabled VMs with kernel version 5.13.  Each laptop is equipped with an Intel Core i7 processor, 16 GB of RAM, and SSD storage. To separate the laptops, a TL-SG108E switch was used with networking interfaces implemented at 1000 Mbps link rates using standard Cat-5 ethernet cables.

\subsection{Test Scenarios and Configurations}

We begin each experiment with pre-distributed IDS modules running within the host IDS agents.  Additionally, each host has an established control channel with the IDS orchestrator.  We now describe each attack and the configurations used for the experiment. 

\subsubsection{DNS Flood Attack}
DNS flood attacks can occur with varying levels of hosts participating in the attack on the network.  Therefore, we assign VM2 as an aggressive attacker, VM3 as mid-level attacker, and VM1 as a benign host for this experiment. Orchestrator response is tested in two iterations with a comparison of pre-built vs.\ JIT module response times.

In the first experiment, each host includes an IDS module to monitor for DNS queries that exceed a predefined frequency limit of 10 requests per second. 
To start the attack on the host, we launch a Python script to query the local DNS server at a rate exceeding the threshold. Upon detection of this attack, the IDS module generates an alert and sends it to the orchestrator.  The IDS orchestrator then builds a new IDS module designed to throttle DNS queries to 5 requests per second. Since a DNS flood attack may come from multiple hosts on the network, the orchestrator is configured to deploy this new module to all hosts on the network, regardless of alert status.
If a host exceeds the threshold after this mitigation, the IDS module 
throttles the DNS query rate to the allowed threshold, without disconnecting the host or killing the offending process.  This differentiation showcases the versatility of IDS modules, highlighting the ability to limit a host without operator interaction.

\subsubsection{Malicious Root Process}

In the second experiment, the host includes an IDS module to monitor for HTTP connections to a known malicious URL.
The second attack begins when the root user makes an HTTP request to the known malicious URL, which generates an IDS module alert that is sent to the orchestrator. In addition to generating the alert, the host module is configured to prevent the connection by not allowing any data to be copied into the socket buffer, effectively nullifying the HTTP request.  The orchestrator processes the alert, selects the appropriate IDS module to increase network defense, and deploys the module to the affected host.  
This new module monitors for any root process that attempts to establish an HTTP connection and, if detected, the connection is blocked and the URL is fed back to the orchestrator to be used in future response modules. 

\subsection{Performance Metrics}\label{test-obj}
Our experimentation focuses on demonstrating the capabilities listed in Section \ref{concept}. Multi-fidelity flow interrogation is demonstrated when an alert detected by a single IDS agent triggers network-wide measures to stop or mitigate a DNS flood attack. The DNS flood experiment measures the time between IDS module alert, orchestration response, and subsequent IDS module attack termination.  We measure this time for both pre-built and JIT module orchestrator response to provide an upper bound since we did not optimize the orchestrator-host communications within the prototype implementation; instead focusing on the interaction.
Experiment two illustrates network-wide event correlation by feeding back malicious request URLs to the IDS orchestrator, which then iteratively builds IDS modules based on network-wide observations.
\section{Experimental Results}

In this section we present the results of the orchestration response to two DNS flood attack scenarios and to a root process attempting to reach a malicious web server.  

\subsection{DNS Flood Attack}

In the first round of the experiment we utilized the JIT module creation method, which required the IDS orchestrator to build response modules after the alert condition is received.  We then repeated the experiment utilizing pre-built attack response modules to show a reduced orchestrator response time is possible.

\subsubsection{JIT Response Modules}
\begin{figure}[htb]
    \centering
    \resizebox{86mm}{!}
        {\includegraphics[width=0.95\linewidth, trim=1 2 1 4, clip]{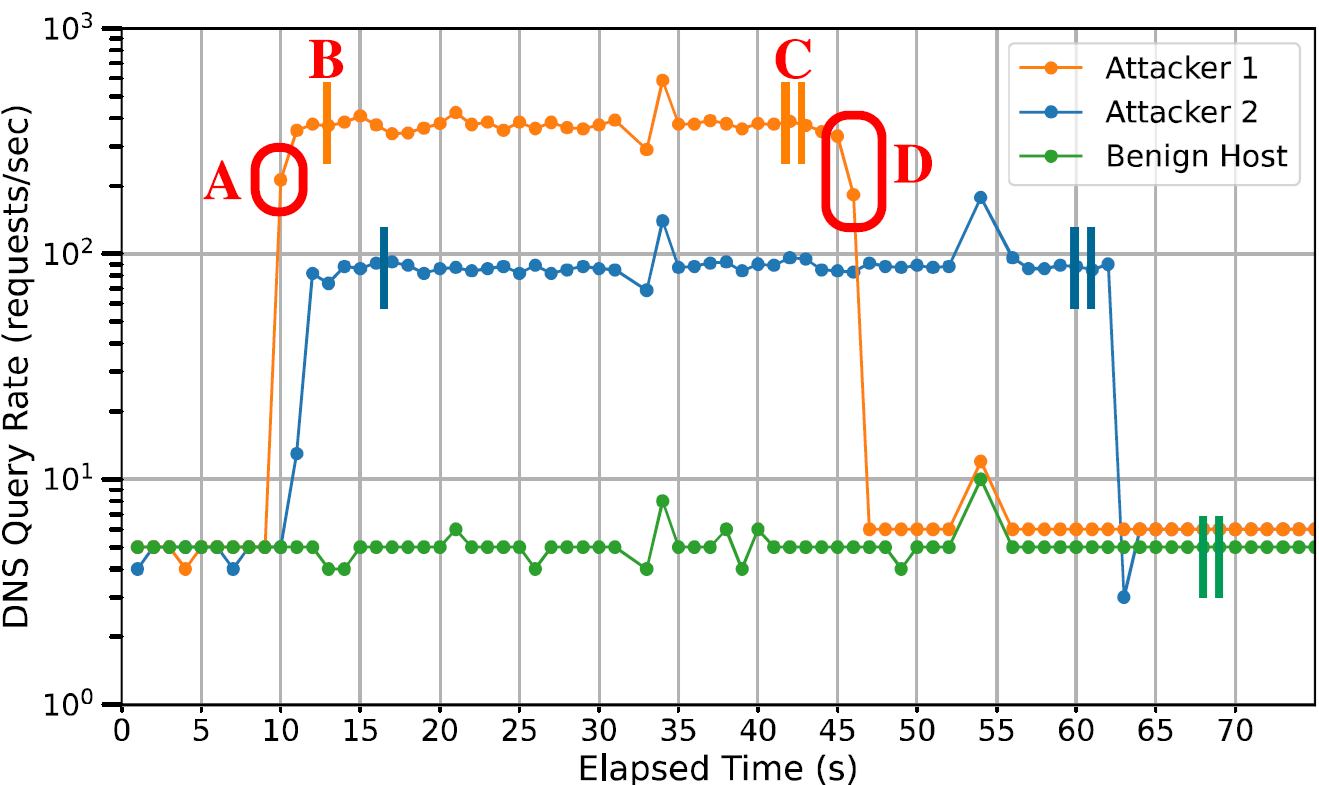}}
        
    \caption{Illustration of mitigation of DNS flood attack using JIT modules. Major events: initial alert from IDS module (\textcolor{red}{A}), orchestrator notification (\textcolor{red}{B}), response module built and deployed (\textcolor{red}{C}), and response effective (\textcolor{red}{D}). For brevity, we only annotate the Attacker 1 events, while using the same style of vertical markers for Attacker 2 and the Benign host.}
    \label{fig: DNS benign}
\end{figure}

Figure \ref{fig: DNS benign} shows the result of the DNS flood scenario using JIT response modules in which the hosts have different attack intensity.
For brevity, we only annotate the Attacker 1 events, while using the same style of vertical markers for Attacker 2 and the Benign host.

At the beginning of the test window, each host starts sending 5 DNS queries per second for the first 10 seconds.  After this normal behavior period, the aggressive and mid-level attackers begin the DNS flood attack (A), exceeding the DNS query rate limit set by the IDS module, which results in an alert being sent to the orchestrator (B). 
The orchestrator alert processor method then determines the next-tier response IDS module, builds the module, and deploys it to all hosts (C).   
The host IDS agent installs the new module (D), which sets the DNS query rate limit to 5 requests per second.  This rate is exceeded almost immediately, resulting in the host being throttled to a maximum of 5 requests per second.  
This experiment demonstrates the discrete ability of network-wide orchestration to apply attack mitigation to select hosts, regardless of current alert status, as seen by the benign host also getting the IDS response module approximately 25 seconds after Attacker 1. 

As discussed in Section \ref{limitations}, the orchestrator monitor window was set to 5 seconds, which minimized the delay between alert generation (A) and orchestrator notification (B).  The majority of response time was the result of JIT module constructions and deployment (C).  One method to reduce this delay would be to utilize pre-built IDS response modules.

\subsubsection{Pre-Built IDS Response Modules}
\begin{figure}[htb]
    \centering
    \resizebox{86mm}{!}{
         \includegraphics[width=0.95\linewidth, trim=1 0 1 4, clip]{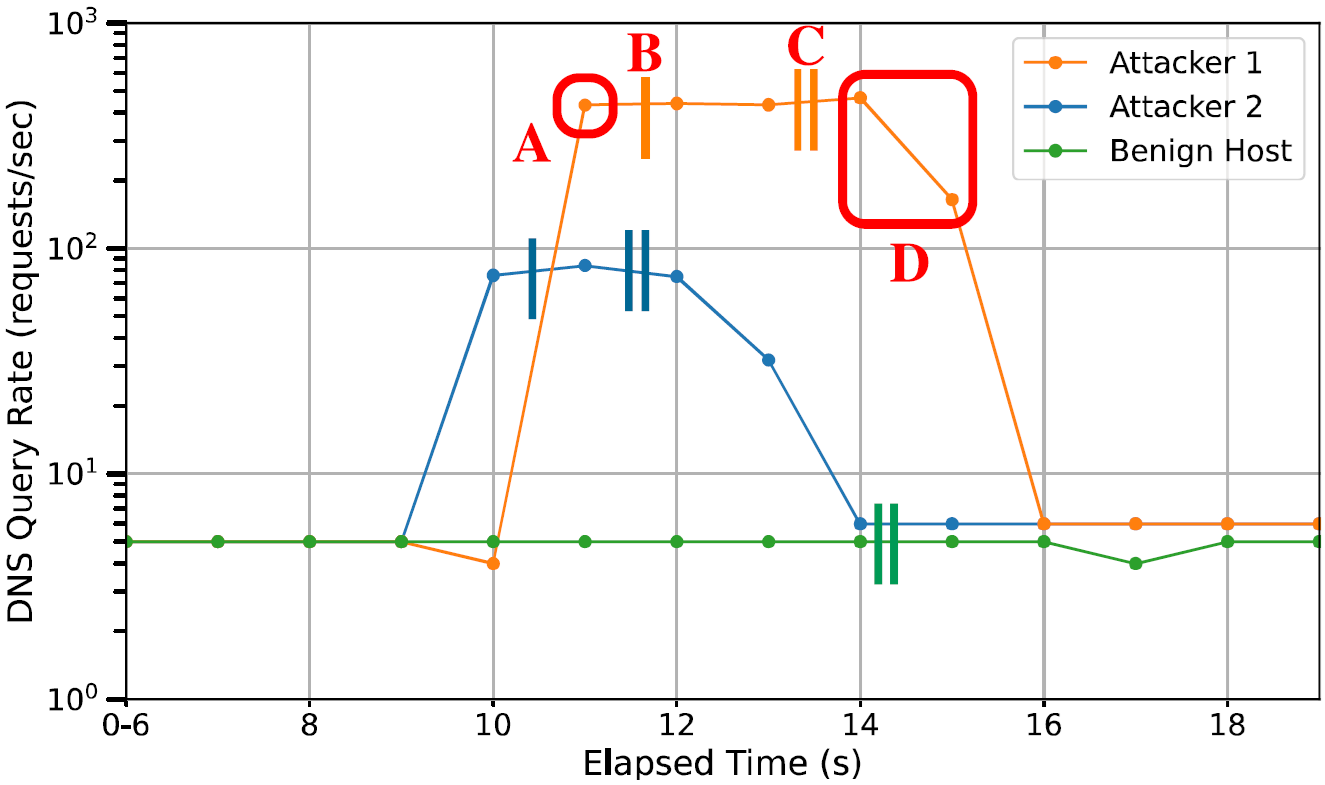}
    }
    
    \caption{Illustration of pre-built IDS module response to DNS flood attack. Major events: initial alert from IDS module  (\textcolor{red}{A}), orchestrator notification (\textcolor{red}{B}), response module deployed (\textcolor{red}{C}), and response effective (\textcolor{red}{D}). Compared to the JIT response (Figure~\ref{fig: DNS benign}), the mitigation time was much shorter. }
    \label{fig: DNS jit}
\end{figure}

Figure \ref{fig: DNS jit} demonstrates how pre-built responses, when applicable, can be utilized to reduce the orchestrator response time.
This approach drastically reduces the response time by building response modules prior to an attack, avoiding the build time of the IDS module upon initial alert detection.  

The overall response time is reduced from more than 30 seconds in the JIT scenario, to less than 10 seconds.  Decreased response time is crucial to defeating network threats promptly, however,  pre-built modules lack the ability incorporate real-time information.  If pre-built modules require new information or logic, they must follow the JIT module methodology.

\subsection{Malicious Root Behavior}
\begin{figure}[htb]
    \centering
    \resizebox{67mm}{!}
        {\includegraphics[width=0.75\linewidth, trim=1 4 1 4, clip]{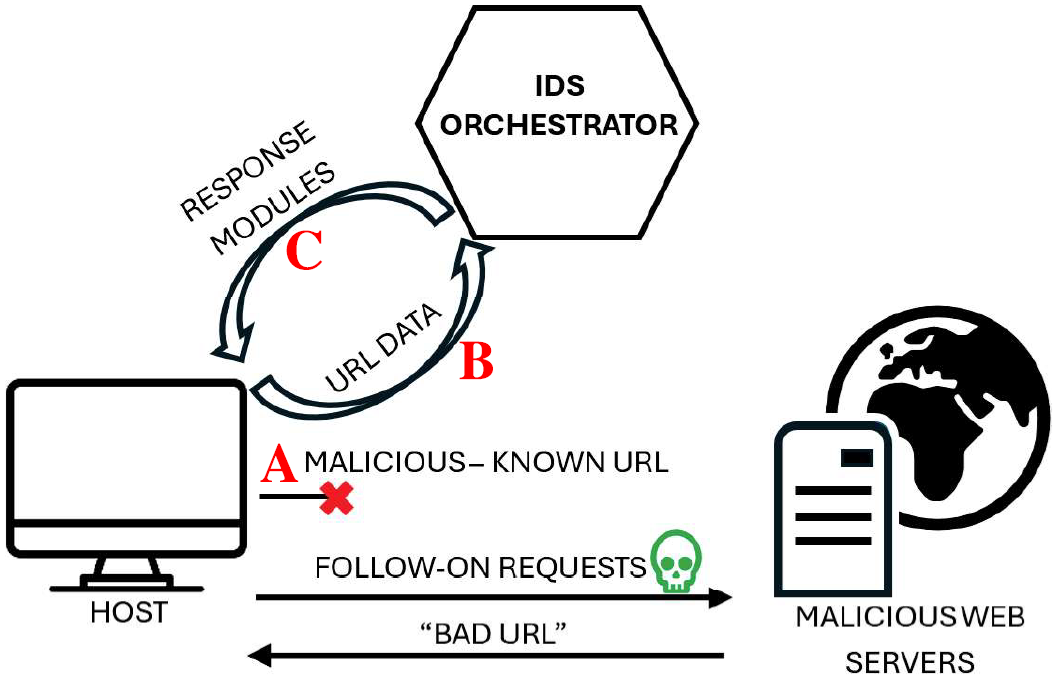}}
    
    \caption{Illustration of mitigation of malicious URL connections. An HTTP request (\textcolor{red}{A}) from a root user process to known malicious URL is blocked, triggering an alert to the orchestrator (\textcolor{red}{B}). Orchestrator deploys a response module (\textcolor{red}{C}). The response module poisons suspicious follow-on requests to prevent connection and continues to feed offending URL data to the orchestrator.}
    \label{fig: HTTP Request}
\end{figure}

This experiment demonstrates IDS module ability to detect malicious root user web access attempts, with subsequent IDS orchestrator response. Figure \ref{fig: HTTP Request} illustrates the attack sequence and orchestrator response.

The attack begins when the root user attempts a web connection (A) to a known malicious URL (``exampleurl.com"). One key difference between this experiment and the previous DNS flood attack is that the initial IDS module is configured to take action against the threat immediately and generate an alert to the IDS orchestrator.  Specifically, if a root process attempts to connect to this URL, the HTTP request is prevented from copying into the socket buffer.  The alert is received by the IDS orchestrator, which deploys a response module that logs all future HTTP requests made by root processes, and poisons the requests to prevent connection establishment.  The logged data is fed back to the orchestrator (B) for use in building the next iteration of response modules (C).  This feedback loop approach allows automatic response to detect, respond, and block malicious processes attempting to connect to previously unseen malicious URLs.

\section{Discussion}

In this section, we discuss two additional design considerations for extending our system with the goal to further evaluate the power and limitations of network-wide IDS orchestration.

\subsection{Top Down Security Enforcement}

We believe the IDS orchestrator can also function as a top-down security policy enforcement point for the entire network.  
As the threat level changes, the orchestrator can proactively adapt the security posture for all hosts, e.g., lock down a port or suspend a set of user accounts, without need for manual reconfiguration. 

We envision network-wide security policy being expressed in formal declarative rules, which are then mapped into host-level actions carried out by low-level security modules that are dynamically deployed to network hosts. 
Our experimentation showcased the ability to vary the level of host scrutiny based on the observed or anticipated network threats.  Future work towards the declarative policy mapping to the security module deployment is required.

\subsection{Insider Threat Detection}

Insider threats are a major concern within any network and can be difficult to detect using methods such as anomaly detection or signatures~\cite{insider}. One reason insider threat detection is difficult on the network is that user traffic is identified by the host IP address instead of the user-name making the request.  Insider threats may also hide within the network by spoofing traffic (e.g., spoofed MAC, IP address, or even user account), which makes detection more challenging.

We believe it would be possible to utilize the central IDS orchestrator to assist with insider threat mitigation. To accomplish user specific IDS modules, future work should integrate the IDS orchestrator with network directory access and authentication servers.  This would allow the orchestrator to build IDS modules to match users throughout the entire network regardless of which machine they utilize and correlate traffic from different user accounts to detect anomalous behaviors.
\section{Related Work}
We referenced related works throughout the paper.
In the following, we briefly discuss two high level concepts, from related work, that have shaped our research.

A defining characteristic of our design is logically centralized network control. This style of control has been designed, adopted, and proven advantageous in different system contexts, including software defined networking (SDN)~\cite{onos}, which is widely deployed in data centers and 5G networks. 
A recent study~\cite{lukaszewski2023AgileNetOps} investigated the specific question of whether continuous, centrally orchestrated monitoring and customization of host behaviors is beneficial, and the results show that such orchestration can increase the agility of enterprise networks, enabling rapid responses to security and performance-related events.

Broadly speaking, our work is an investigation of autonomous cyber defense. Prior efforts have explored other aspects of autonomous cyber defense that are complementary to our focus, including the deployment of AI agents~\cite{gabirondo21} and the use of deep reinforcement learning~\cite{nguyen21}. 
\section{Conclusion}

This work reinforces the utility of integrating host-based intrusion detection with centralized orchestration to create a more adaptive and resilient cybersecurity framework. We presented a system that leverages socket-level monitoring and dynamic IDS module deployment to mitigate threats at the host level while simultaneously influencing network-wide security policies. While we demonstrated a limited set of functionality with our prototype, we believe that the approach of network-wide orchestration of host IDS actions has the power to  enhance situational awareness, reduce false positives, and ultimately, enable automated defensive measures that adapt in real time to evolving threats with minimum human interventions. 

\bibliographystyle{abbrv}

\end{document}